\documentstyle[twoside,fleqn,espcrc2,psfig]{article}

\newcommand{\al}{\alpha}
\newcommand{\bt}{\beta}

\newcommand{\be}{\begin{equation}}
\newcommand{\ee}{\end{equation}}
\newcommand{\ba}{\begin{eqnarray}}
\newcommand{\ea}{\end{eqnarray}}

\newcommand{\dd}{\partial}

\newcommand{\vfi}{\varphi}

\newcommand{\AmS}{{\protect\the\textfont2
  A\kern-.1667em\lower.5ex\hbox{M}\kern-.125emS}}

\title{One-particle Hilbertspace of 2+1 dimensional gravity using
       non-commuting coordinates}

\author{M. Welling\address{Institute for Theoretical Physics,
                           University of Utrecht, \\
        P.O. Box 80006, 3508 TA Utrecht, The Netherlands}%
        \thanks{Work supported by the European Commission TMR programme
                ERBFMRX-CT96-0045}}

\begin{document}

\begin{abstract}
After a review of multi-particle solutions in classical 2+1 dimensional gravity
we will construct a one-particle Hilbertspace. As we will use a curved momentum
space, the coordinates $x^\mu$ are represented as non-commuting Hermitian
operators on this Hilbertspace. Finally we will indicate how to construct a
Schr\"odinger equation.
\end{abstract}

\maketitle

\section{INTRODUCTION}
In 1963 Staruszkiewicz \cite{eerst} considered general relativity in 2+1
dimensions for the first time. He solved the gravitational field surrounding a
point-particle and found that is represents a conical spacetime. The
subject was revived in 1984 by Deser, Jackiw and 't Hooft \cite{beginarticle}
where it was shown that multiparticle solutions can be constructed by cutting
wedges out of spacetime and identifying the boundaries according to a
Poincar\'e
transformation. This idea was worked out by 't Hooft in \cite{tHooft1}
who also proved that
for closed universes there can be no closed timelike curves.
It is important to notice that one will never find gravitational
wave solutions in 2+1 dimensions because the gravitational field carries no
degrees of freedom. So all degrees of freedom must come from either topology
(handles) or from matter.

A completely different viewpoint on the subject was
put forward by Achucarro and Townsend (1986) \cite{CS1} and Witten (1988)
\cite{CS2}. They proved that 2+1 dimensional gravity was equivalent to a
Chern-Simons theory
with the Poincar\'e group as its gauge group.

Quantization programs have mainly
concentrated on matter free universes with torus or higher genus topology
\cite{Carlip}.
In this paper we will treat the quantization of one particle states
as was advocated by 't Hooft \cite{tHooft2} and possible variations on that
theme.

\section{CLASSICAL MULTI-PARTICLE SOLUTIONS}

\begin{figure}[htb]
\vspace{9pt}
\psfig{figure=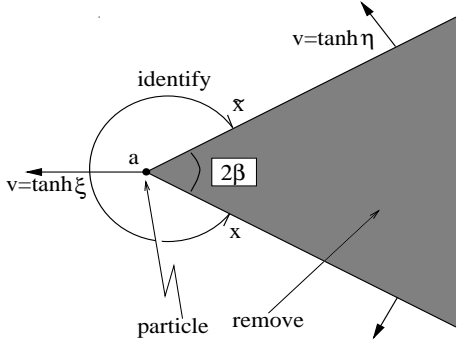,height=4.5cm,width=6cm,angle=-90}
\caption{Conical space surrounding a moving particle.}
\label{one-particle}
\end{figure}

If we solve the gravitational field surrounding a static point particle
we find that it is a conical space. Therefore we may choose flat
(Minkowskian) coordinates globally but with unconventional ranges. In polar
coordinates ($r=0$ is the position of the particle), the angle $\vfi$ runs
from $0$ to $2\pi(1-4Gm)$, where $G$ is Newton's constant and $m$ is the mass
of the particle. So we can picture space by cutting out a wedge and identifying
the boundaries. To describe a moving particle
we simply boost this solution. The Lorentz contraction widens the angle of the
wedge that is `missing' from spacetime. Also the identification rule is now
a Poincar\'e transformation of the following form:
\be
{\bf x'}={\bf a}+BRB^{-1}({\bf x}-{\bf a})
\ee
Here ${\bf x}$ and ${\bf x'}$ are opposite points on the boundaries, ${\bf a}$
is the position of the particle, $B$ is a boost matrix and $R$ is a rotation
over an angle $8\pi Gm$.
We have pictured the situation in figure (\ref{one-particle}) where one can
also
find the variables $\bt$ (half the deficit angle), $\eta$ (perpendicular
rapidity of
the boundary), $\xi$ (the rapidity of the particle: $v=\tanh(\xi))$ and
$\mu=4\pi Gm$.
It is easy to deduce some relations among these variables:
\ba
\tan(\bt)&=&\cosh(\xi)\tan(\mu)\\
\tanh(\eta)&=&\sin(\bt)\tanh(\xi)\\
\cos(\mu)&=&\cos(\bt)\cosh(\eta)\label{sv}\\
\sinh(\eta)&=&\sin(\mu)\sinh(\xi)
\ea

It is important to notice that we chose the wedge `behind' the particle so
that we can avoid time jumps across the wedge.

\begin{figure}[htb]
\vspace{5pt}
\psfig{figure=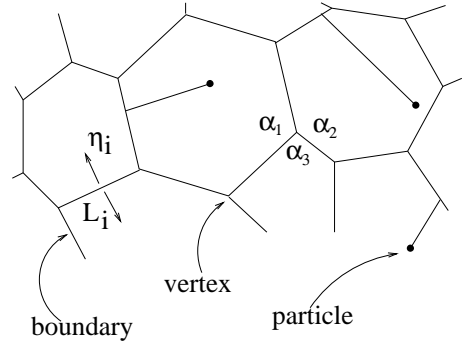,height=4.5cm,width=6 cm,angle=-90}
\caption{A Cauchy surface containing particles made of polygons.}
\label{polygons}
\end{figure}

If we want to describe a multiparticle soltion it is convenient to construct
a cauchy surface consisting of flat patches of Minkowski space. The patches
must be glued together in such a way that the metric is continuous across
the boundaries. The result of such a construction is that the boundaries as
seen by observers on the neighbouring patches has equal length and can only
move  perpendicular to itself. Moreover the velocities in both frames have the
same
magnitude, but not necessarily the same sign. On the spot where three edges
meet we have a vertex. The angles $\al_i$ (see figure (\ref{polygons})) need
not add up to $2\pi$ so that we can construct curved surfaces.

We can
introduce particles on this surface by putting them inside a polygon. The
`tailpipe' introduced earlier connects to a neighbouring patch where it forms
a vertex. Of course, besides the vertices, also particles introduce curvature
on the surface. It should be noted however that the three dimensional curvature
of a vertex vanishes as there is no matter present in contrast to the
three-curvature of a particle which is proportional to its mass. If the system
evolves in time all edges shrink or grow linearly in time. So now and then an
edge appears or disappears (according to certain rules that can be derived).
It is also possible that a particle hits a boundary of a patch after which it
proceeds with a Lorentztransformed velocity in an other polygon. We will call
these events `transitions'.
It is  important that this is a completely deterministic system that evolves
Cauchy
surfaces in time. This is why there will be no trouble with causality. It is
even
possible to reformulate it as a Hamiltonian system. If we choose our
Hamiltonian to be the total deficit angle of the surface (which equals the
total energy
contained in that surface) and the length $L_i$ as our configuration variables
then we find (by solving Hamilton's equations) that the momentum conjugated to
the boundary variables is $p_i=2\eta_i$ which is the rapidity with which the
boundary moves. One may notice now that there are far too many degrees of
freedom, as the only degrees of freedom are connected with the particles and
there are many more boundaries. This is due to the fact that there are also
some constraints in the model connected with the closure of the polygons. For
instance, the angles inside a polygon (which are functions of the momenta)
should add up to $(N-2)\pi$ ($N$ is the number of edges surrounding a polygon).
There are also two more constaints to ensure that
the last boundary $L_N$ of a polygon, viewed a vector for a moment, bites the
first boundary $L_1$ in its tail. These constraints constitute a system of
first
class constraints which generate `gauge-transformations' in the following
sense.
The closure of angles generate time translations of that particular polygon,
and the closure of boundaries generate Lorentztransformations of the polygon.
So we have now a constrained Hamiltonian model at our disposal for which
quantization seems straightforward. The transitions however, that must be taken
into account as boundary conditions on the wavefunction, are troublesome. This
is the reason why we will first concentrate on the one particle quantization.

\section{ONE-PARTICLE HILBERT SPACE}

In this section we will follow a quantization scheme first proposed by Snyder
in 1947 (!)
\cite{Snyder} and reinvented by 't Hooft \cite{tHooft2} in the case of 2+1
gravity. The main idea behind Snyders paper was to introduce a curved momentum
space (he used De Sitter space) that still has the maximal number of symmetries
among which the full Lorentzgroup.
In the case of De Sitter space or anti-De Sitter space one trades the
translations of the
Poincar\'e group for four more (in the case of 3+1 dimensions) Lorentz type
transformations. So one cannot expect to preserve the full group of
translations as an
invariance group. But the amazing thing is that we can still define coordinates
as Hermitian operators that act on wavefunctions living on this curved (but
maximally symmetric) momentumspace that transform covariantly under the full
group of Lorentztransformations. In the previous section we have seen that the
variables $L_i$ and $2\eta_i$ are each others conjugate and that
$\eta_i$ is really a hyperbolic angle. For one particle there is an alternative
pair of conjugate variables. If we denote the position of the particle in the
two dimensional plane by $(x,0)$ and give a speed $v=\tanh(\xi)$ in the
$x$-direction than we find that the conjugate momentum is an angle $\theta$.
The `Schr\"odinger equation' (\ref{sv}) then becomes \cite{tHooft2}:
\be
\cos(H)=\cos(\mu)\cos(\theta) \label{sv2}
\ee
Moreover, we have seen in section 2 that the Hamiltonian was also given by an
angle. The next step is to choose a curved momentum space. There are many
possibilities. Denote by $H^p_q$ a hyperboloid given by the
following relation:
\be
Q_1^2+...+Q_p^2-Q_{p+1}^2-...-Q_q^2=1
\ee
Inspired by the fact that the momentumvariables and the Hamiltonian are given
by angles
't Hooft studied the possibilities:
\begin{itemize}
\item[a)]  $H^3_0\times H^2_0$ ~~~~(=$S^2\times S^1$)
\item[b)]  $H^4_0$ ~~~~(=$S^3$)
\end{itemize}
So in the first case the momentum variables live on a sphere and the energy
lives on a
circel, in the second case all variables are combined in a three sphere.
But there are also different possible choices which are presently being studied
by the author. Interesting choices seem to be:
\begin{itemize}
\item[c)] $H^2_1\times H^2_0$
\item[d)] $H^2_2$
\end{itemize}
How can we do quantummechanics on these spaces? We should of course study
wavefunctions that live on these homogeneous momentum spaces. In particular we
would like a complete set of orthonormal, square integrable functions to define
a basis in our Hilbert space.
Fortunately there is a lot of literature on this subject and one of the results
is that on all this homogeneous spaces there exists such a complete,
orthonormal, square integrable (with respect to a suitable measure) set of
basisfunctions.
For instance, on the sphere we have the well known spherical harmonics $Y_{\ell
m}(\theta,\varphi)$ as our basis.

Let us elaborate a bit on case a). The sphere is given by the equation:
\be
Q_1^2+Q_2^2+Q_3^2=1
\ee
If we define:
\ba
x^k&=&i\ell_P(Q_3\frac{\dd}{\dd Q_k}-Q_k\frac{\dd}{\dd Q_3})\\
t&=&i\ell_P\frac{\dd}{\dd H}\\
\ea
and
\be
\tan(\frac{\ell_P}{\cos(\mu)}p_k)\equiv\tan(\theta_k)=\frac{Q_k}{Q_3}
\ee
where $k=x,y$ and $\ell_P$ is the Planck-length,
then the coordinates $x^k$ and the vector $(\tan(\theta_x),\tan(\theta_y))$
will transform covariantly under rotations generated by:
\be
L=i(Q_2\frac{\dd}{\dd Q_1}-Q_1\frac{\dd}{\dd Q_2})
\ee
This is of course checked by calculating the commutators:
\ba
&&{[L,x]}=-y~~~[L,\tan(\theta_x)]=-\tan(\theta_y) \\
&&{[L,y]}=x~~~~~~[L,\tan(\theta_y)]=\tan(\theta_x)
\ea
The price that we are paying is that the usual commutation relations among the
phase space variables are changed. For instance:
\be
{[x,y]}=i\frac{\ell_P^2}{\cos^2(\mu)}L
\ee
This implies that also the coordinates become subject to uncertainty relations.
If we choose topology d) as our momentum space, also the time coordinate mixes
into the non-commutative structure. In that case Lorentz transformations become
simple pointtransformations.

Finally we would like to comment on the construction of a Schr\"odinger
equation on these spaces. The basisfunctions of the momentum space are
polynomials of the embedding coordinates $Q_\nu$. For instance in the above
example the basisfunctions are:
\be
\Psi_{\ell,m,t}(Q_j,H)=Y_{\ell m}(Q_j)\exp[iHt]
\ee
where $j=1,2,3$ and $\ell$ is the degree of this polynomial. The equation
(\ref{sv2}) can be written in terms of the $Q$-variables:
\be
\cos(H)\Psi_{l,m,t}(Q_j,H)=\cos(\mu)Q_3\Psi_{l,m,t}(Q_j,H)\label{sv3}
\ee
The action of $\cos(H)$ on $\Psi$ is simple: it is a shift of one time step in
the possitive direction minus a shift of one time step in the negative
direction. Because $Q_3=Y_{10}$, the action of $Q_3$ on $\Psi$ is just the
calculation of Clebsch-Gordon coefficients for the decomposition of the tensor
product of two SO(3) representations in its irreducible representations. In
this case we find that the action of $Q_3$ is a linear combination of shifts of
$\ell$ of one step in the positive and negative direction. Because of the time
steps in positive {\em and } negative direction, equation (\ref{sv3}) is the
analogue of the Klein-Gordon equation and therefore suffers from the same
disease: $|\Psi|^2$ cannot be interpreted as a probability distribution. The
solution for this problem is to construct a Dirac-like equation (see
\cite{tHooft2}).

When we use topology d) as our momentum space, the Hilbertspace is given by
polynomials of the $Q_\nu$ where $\nu$ runs from 1 to 4. They are the infinite
dimensional representations of SO(2,2) (which spectrum contains a continuous
and a discreet part) and form an orthonormal set of square integrable
basisfunctions. Although the $Q_\nu$ transform now under the nonunitairy finite
dimensional representations of SO(2,2) their action on the infinite dimensional
unitairy representations can still be calculated using recurrence relations of
the hypergeometric function. Although the Dirac-like equation is not completely
satisfactory yet the advantage of this Hilbertspace seems to be that
Lorentztransformations act very easy on it; they only involve nearest
neighbours.

For two particles the complicated boundary conditions on the wave functions
play an essential role. To formulate these boundary conditions on the wave
functions it is very convenient to have a discreet spacetime. The above
quantization procedure seems to provide that structure.

\end{document}